# Using Agent to Coordinate Web Services

C. H. Liu, Y. F. Lin and Jason J. Y. Chen
Department of Computer Science & Information Engineering
National Central University
Jhong-Li, Taiwan
jasonjychen@gmail.com

*Abstract*— **Traditionally, agent and web service are two separate research areas. We figure that, through agent communication, agent is suitable to coordinate web services. However, there exist agent communication problems due to the lack of uniform, cross-platform vocabulary. Fortunately, ontology defines a vocabulary. We thus propose a new agent communication layer and present the web ontology language (OWL)-based operational ontologies that provides a declarative description. It can be accessed by various engines to facilitate agent communication. Further, in our operational ontologies, we define the mental attitudes of agents that can be shared among other agents. Our architecture enhanced the 3APL agent platform, and it is implemented as an agent communication framework. Finally, we extended the framework to be compatible with the web ontology language for service (OWL-S), and then develop a movie recommendation system with four OWL-S semantic web services on the framework. The benefits of this work are: 1) dynamic web service coordination, 2) ontological reasoning through uniform representation, namely, the declarative description, and 3) easy reuse and extension of both ontology and engine through extending ontology.**

## I. INTRODUCTION

Traditionally, agent and web service are two separate research areas. The research on agent focuses on problem solving mechanisms in distributed environment. A widely known agent model is the belief, desire and intention (BDI) model. Each agent has its capability (actions). Also, each has its mental attitude (what it believes). On the other hand, the research on web service concentrates on distributed technique and standard such as web service description language (WSDL), simple object access protocol (SOAP) and universal definition and discovery integrated (UDDI). However, web services are not reliable and easy-to-use due to the fact that they are at remote sites that a user has no control over them. Additionally, the web services provide a static description on the Web, thereby making it more difficult for users to use them.

Incidentally, the Semantic Web is a highly-anticipated infrastructure for agents to run on it and to perform complex actions for their users [1] [2] [3]. Further, an agent is suitable to coordinate with each other with the same purpose in distributed environments [4] [5]. For example, an agent is able to proactively request other agents for a movie recommendation, and then the requested agents respond reactively the name of some movies such as "Night at the museum" or "Brokeback Mountain" through agent communication. For agent communication, the Foundation for Intelligent Physical Agents (FIPA) interaction protocol forms a suite of protocol standards, which defines 22 communicative acts. Through the communication mechanisms, the agent decides the proper agents to execute their actions, which are implemented as semantic web services that in turn are annotated web services in this paper. Thus, the web services can be coordinated through the agent communication [6].

A problem with agent communication is that there is no cross-platform, uniform vocabulary to identify concepts used in various agent programs. Further, when an agent communicates with other agents, it cannot predict the mental attitudes of other agents with its mental model. This causes difficulty and complexity in communication. The method of agent communication is thus questioned by some researches [7] [8]. John Yen [9] introduced a manner in which the agents can form a team and share mental models with each other in order to make decisions. Fortunately, ontology is a document that formally defines a vocabulary [10]. The web ontology language (OWL) is a popular language to describe ontology, and it allows people to utilize the uniform referential identifier (URI) to give every concept a specific term such as those used in agent communication. It also defines semantics of the terms, and organizes all kinds of terms by using relations among the terms. The terms can be shared among agents.

A special ontology is called operational ontology, in which the operational concepts are represented in a declarative description. By using OWL to describe the operational ontology [11], this problem can be solved. Further, the mental attitudes of agents are shared among other agents, which is called proposition ontology (to be covered shortly).

This paper presents an agent communication framework with operational ontology, as well as the mental attitude of agent, to facilitate agent communication.

This paper is organized as follows. Firstly, we depict the agent communication layers stack and the ontology associated with each layer in section 2. Secondly, we explain the agent communication architecture that consists of the declarative descriptions for ontology and the engines to access them. Next, the architecture is implemented to a framework by enhancing the 3APL agent platform in section 3. Then, the framework is extended to be compatible with the web ontology language for service (OWL-S). Further, we develop a movie recommendation system with four OWL-S services on our framework in section 4. Lastly we give the conclusion in section 5.

## II. AGENT COMMUNICATION LAYER AND ITS ONTOLOGY

We propose the new agent communication layer with operational ontologies. Firstly, we will introduce the new agent communication layer in section 2.1, and then describe its ontology in section 2.2.

### A. New agent communication layer

The new agent communication layer, called the Agent Mentality Layer, lie between the Content Language Layer and the Message Transport Layer. It makes the message transported capable of transmitting mental attitudes of agents. Thus, the mental attitudes could be shared among agents. Ontologies are vocabularies that should be obeyed when agents are communicating with each other. For each agent communication layer in Figure 1, a corresponding operational ontology is also defined.

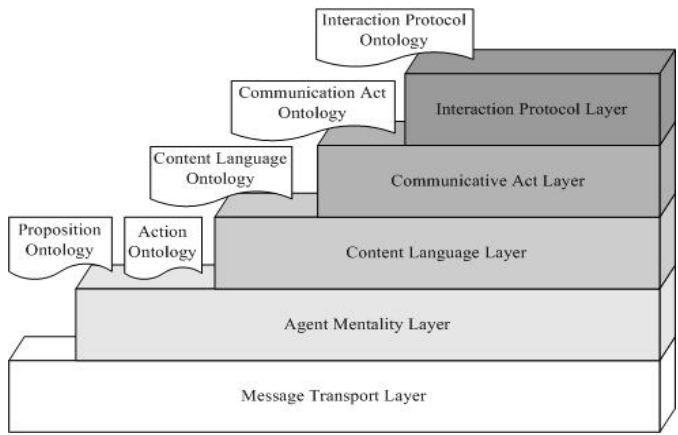

Figure 1: A New Agent Communication Layer Stacks

Note that the message transport layer is not covered in the paper. In the agent mentality layer, the proposition ontology is used to define the representation of mental attitudes of agents, which is the abstract concept of agents and is represented by property "belief". Also, in this layer, action and capability of agents are defined by action ontology. In the content language layer, the content language ontology defines the message representation. In the communication act layer, communicative act ontology defines formal semantics of every communicative act. As for the highest interaction protocol layer, the interaction protocol ontology defines the state shift during communication. It constitutes all communication acts and thus forms a conversation pattern, which then becomes a restricted standard of communication amid agents. Briefly summing up, the five operational ontologies are collectively called the agent communication ontology, and the documents of these ontologies are referred to as the agent communication description.

The relationship of the documents is shown in Figure 2. Every description in Figure 2 refers to one another via a URI reference. Every interaction protocol description refers to multiple communicative act descriptions. Similarly, every communicative act refers to an action description as well as a proposition description in the content language description.

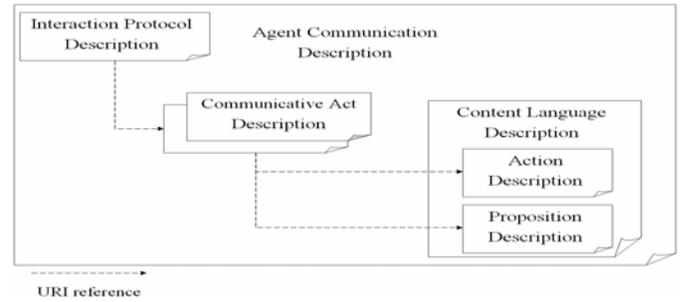

Figure 2: Agent Communication Description

### B. Agent communication ontology

The agent communication ontology provides the formal specifications of the terms and their relations in the domain of agent communication. In the following, we will briefly introduce three main parts of agent communication ontology: (1) Interaction Protocol Ontology, (2) Communication Act Ontology and (3) Proposition Ontology.

Figure 3 illustrates the interaction protocol ontology, which has four main concepts:

- Protocol: It has the constructedBy property, which constructs the interaction protocol, and it also has the hasInitiator and hasParticipant properties, which indicate the agents involved in this interaction.

- Initiator and Participant: An initiator initiates the interaction, and participant(s) represents the agent(s) involved in the interaction.

- State: It represents the interaction status of agents, which has three subclasses: (1) start state, (2) transit state and (3) accept state. Start state is at the beginning of the interaction; the accept state is at the end of the interaction, and the transit state is the intermediate state.

- Transition: It stands for the transition between the state and communication. Through the "execute" property, it could trigger a communicative act.

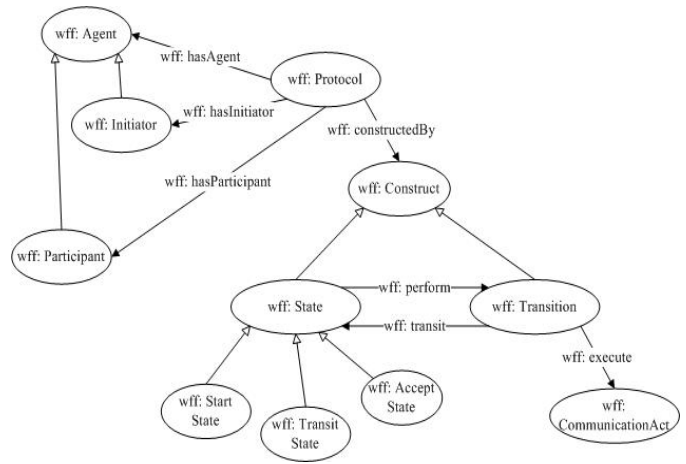

Figure 3: The Interaction Protocol Ontology

Figure 4 illustrates the communication act ontology. In order to define the communicative act ontology, the classification of communicative acts presented here is based on the works of Searle [12]. In our classification, the expressives represent the content, which contains a reason and an action; the commissives represent the content, which contains a condition and an action; the directives represent the content, which only contains an action; and the assertives represent the content, which contains only a proposition.

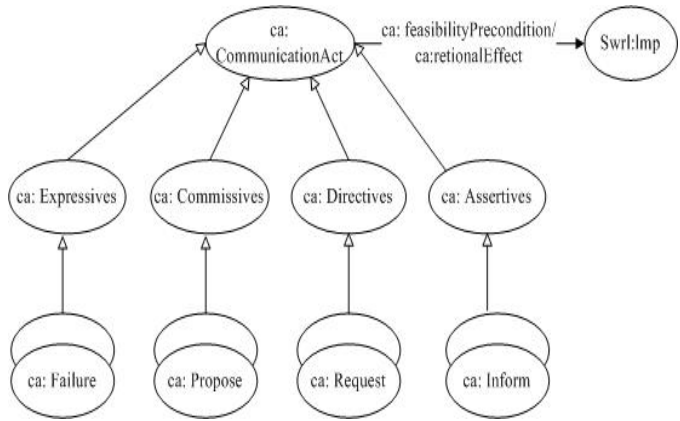

Figure 4: The Communication Act Ontology

Figure 5 shows the proposition ontology, which formalizes the agent mental model shared among agents. It is useful to construct the cooperation among agents. The proposition ontology has three main parts as described below:

- Proposition: A proposition which is described in three parts — subject, predicate and object — and it represents a statement.

- Extended proposition: Extended proposition is a proposition, which adds one valid time to represent the time when the proposition occurs to one agent.

- Embedded proposition: Embedded proposition is an extended proposition, which adds the "belief" to indicate whether someone believes in this proposition or not.

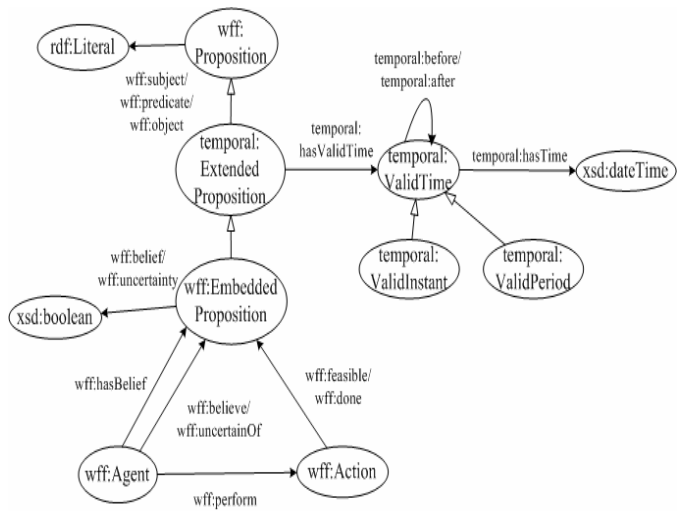

Figure 5: The Proposition Ontology

## III. AGENT COMMUNICATION ARCHITECTURE

As shown in Figure 6, our architecture contains the Mental Model, Action Engine, Proposition Engine, Communicative Act (CA) Engine, and Protocol Engine. Additionally, the architecture also contains 3APL agent platform as the planner, and Data Model as the ontology storage.

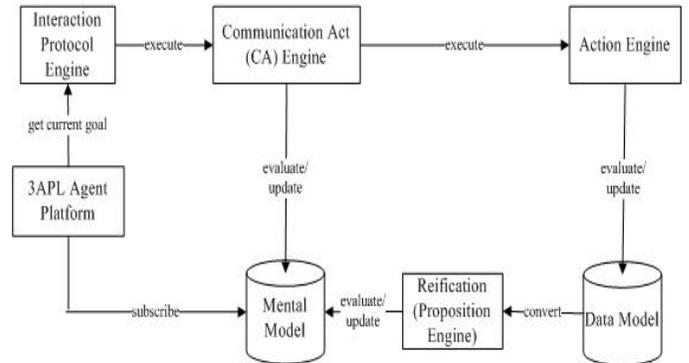

Figure 6: Agent Communication Architecture

Next we will dilate each component in the architecture.

- Interaction Protocol Engine: The interaction protocol engine processes interaction protocol description. It identifies the agent who participates in communication, records the communication state, and then invokes CA Engine to transform the state.

- CA Engine: The CA engine processes communicative act description. It evaluates the feasibility precondition, updates the mental model by rational effect. If action exists in the description, the action engine will be invoked.

- Action Engine: The action engine processes the action description. It decides the way to execute actions such as sequence, concurrence, alternative and iteration. Moreover, the actions' precondition will also be evaluated before execution, and then the actions' effect will be updated in the data model after execution.

- Proposition Engine: The proposition engine processes the proposition description. It reifies a statement in data model as a proposition in mental model. It allows a statement to have different manifestations if needed.

- 3APL Agent Platform: A 3APL is a Prolog-based agent platform. It contains a goal base, a belief base, and a plan base that stores goals, beliefs, and plans, respectively. Besides, it contains plan rules to infer plans and an mProlog-based engine [13] to deliberate these rules. Our work tries to extend this platform with subscribing the proposition in the mental model as data source of belief base.

- Mental Model: The mental model stores the reified statements i.e. propositions. It provides a belief algorithm [14] to appropriately allocate the proposition as an actual world or imaginary worlds.

- Data Model: The data model [15] stores domain ontologies and statements which are generated form

these ontologies. The ontologies are regarded as data types of actions' input/output and generate statements as their values.

## A. Executing agent communication description

Our agent communication is based on the FIPA formal semantics, which is questioned by some researches [16] [17] who point out that an agent is unable to predict mental attitudes of other agents by checking its own attitudes. Thus, if we regard these attitudes as precondition for communication, they should be checked. Further, if we consider them as effect of communication, they probably cannot be determined by the sender agent, because the effect is administered by the recipient agent. Therefore, we would say an agent itself cannot predict an intended purpose.

In our architecture (see Figure7), the mental model is defined by OWL and shared among agents. Further, the data model saves an action's effect and output executed by an agent, which is sharable too. Thus, there is no above mentioned problem.

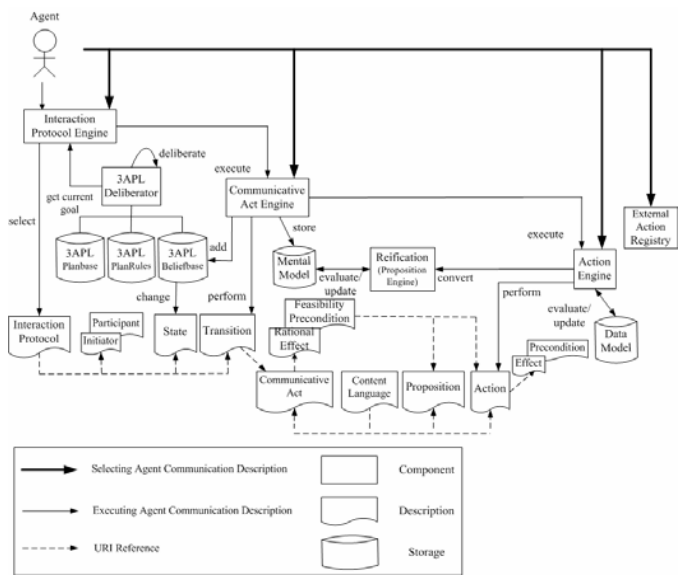

Figure 7: Agent Communication Framework

The execution process of the engines and the declarative descriptions is shown in six steps below:

Step 1: The interaction protocol engine will read the agent communication description. It will read the section of the interaction protocol description and then confirm participative agents by using the initiator and the participant. Afterwards, it will put decomposition rules (plan rules) [18] owned respectively by agents into plan bases of the 3APL agent platform.

Step 2: The interaction protocol engine then enters a certain stage and orders the CA engine to perform the transition. Then it considers the internal communicative act as a parameter, which will be transmitted into the CA engine with the initiator and participant.

Step 3: By viewing the agent mental states in the mental model, the CA engine can evaluate the feasibility precondition (FP) of this CA. If there is no state left, this algorithm terminates. In contrary, It is necessary to meet the demand of the FP and insert all parameters, including the initiator, the participant (come from the protocol description, and they are necessary), the action and the proposition (come from the content language description, and they are not necessary) into it. Then, it will call respectively the action engine and proposition engine to execute processes depicted below. If not, the interaction will come to an end.

Step 4: By viewing the domain knowledge in the data model, the action engine can evaluate the precondition of the action. If it forms, this action will be executed, and the effect of the execution will be recorded in the data model. At the same time, if there has an output which will be transmitted to the propositional engine and be reified to mental model. By contraries, if it fails to form, the execution of the action will be called off. Next, the explanation of this cancellation, such as the precondition and the input bringing about the failure, will be transmitted to the propositional engine and be reified mental model.

Step 5: When the execution of the action is finished, the CA engine can store outputs and reasons that are reified from the proposition engine. Those outputs and reasons are very useful when we are ought to judge the CA next time.

Step 6: Go back to Step 2 and enter the next stage.

## B. Selecting agent communication description

In section 3.1, we described how agents communicate through the agent communication description. Here, we are going to elaborate how agents select appropriate agent communication description before communication. The selection process and the interaction of engines will be discussed next.

If the planner infers plans to be "external actions" (not agent capabilities by itself), the planner knows the agent itself cannot complete a certain task, it will seek help from external agents to move closer to the goal. The agent can coordinate or cooperate with external agents for that. The difference between coordination and cooperation is that if descriptions are protocols [19] [20] of contract net family, the agent uses coordination; inversely, if they are protocols [21] [22] of request family, the agent uses cooperation. Certainly, there are also protocols [23] [24] of broker family. However, that involves interaction of all agents, and it is not discussed here.

The process of selection can be divided into three steps: 1) external action selection and 2) communicative act selection and 3) interaction protocol selection.

Step 1: An external action selection. Agents would query the action registry about the location of the agents with the capability of executing this external action. If they find out an agent, they will send the action description into the action engine; if not, they will terminate.

Action engine evaluates effects in action description with data model as the information source to supply necessary parameters such as output or other local variables. If evaluations succeed, proposition engine will reify and transmit the effects to mental model where they can be preserved. On the contrary, if evaluations fail, it will terminate. Note if this action description was ever executed by agents offering actions, output will remain to exist in data model. Thus, we have to make agents be able to share data model and mental model with each other.

Step 2: A communicative act selection. The CA engine use external actions and output of mental model as parameters to evaluate rational effect of a number of the FIPA 22 CAs descriptions [25]. If evaluations are successful, it will choose and send this CA description to interaction protocol engine; if not, it will go back to the 1.

Step 3: An interaction protocol selection. After the protocol engine receives CA descriptions, with which it will choose the interaction protocol description. Later, the chosen protocol description will replace the original external action. Then, it will be finished.

Our action communication description is the description which linked the action description with the CA description as well as the interaction protocol description.

*C. Aagent communication enchanced 3APL deliberation cycle*

By using agent communication description just mentioned, the 3APL deliberation cycle is enhanced, as shown in Figure 8.

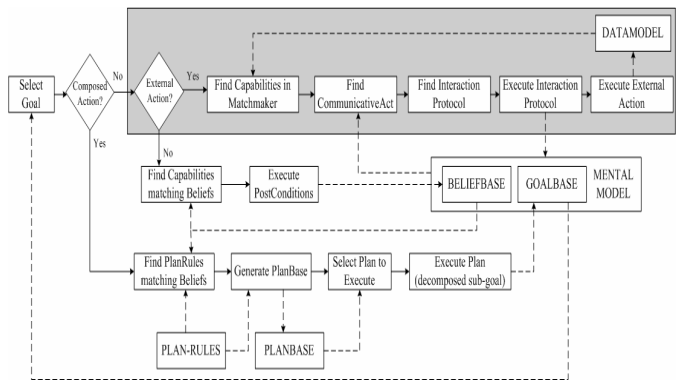

Figure 8: Agent Communication Enhanced 3APL Deliberation Cycle

The agent chooses one goal in goal base, and judge whether it is a composed action or not. If yes, the agent will choose the plan rule to decompose and execute it (the condition of choice is that if the current state of brief case satisfies the guard of the plan rule). After the execution, the newly-produced plan will be stored in the plan base. At the same time, for the purpose of completing sub-goals decomposed by the plan rule, the agent will choose a new plan rule to carry out.

The agent chooses one goal in goal base, and judge whether it is a composed action or not. If yes, the agent will choose the plan rule to decompose and execute it (the condition of choice is that if the current state of brief case satisfies the guard of the plan rule). After the execution, the newly-produced plan will be stored in the plan base. At the same time, for the purpose of completing sub-goals decomposed by the plan rule, the agent will choose a new plan rule to carry out.

If this goal is a basic action, the agent will make a judgment if this action is external or not. If not, the agent will find the capabilities in the belief base. If they are found, this action will be carried out.

If the action is external, it means that the agent itself cannot carry it out. At this point, the agent will question the matchmaker: "Which agent is able to execute this external action?" Subsequently, the agent will decide on one interaction protocol by searching for proper communicative act and follow this protocol to coordinate or collaborate with another agent that will execute this external action.

IV. AGENT COMMUNICATION FRAMEWORK FOR SEMANTIC WEB SERVICES

We extend the framework in Figure 7 to be compatible with the OWL-S, as shown in Figure 9. As mentioned earlier, the action ontology is defined in the agent mentality layer. In Figure 7, we extended the action ontology to refer to the declarative description, which is represented by OWL-S Service description. Then, we can reuse OWL-S's process engine [26] and OWLS/UDDI Matchmaker [27] to execute OWL-S process and OWL-S profile, respectively.

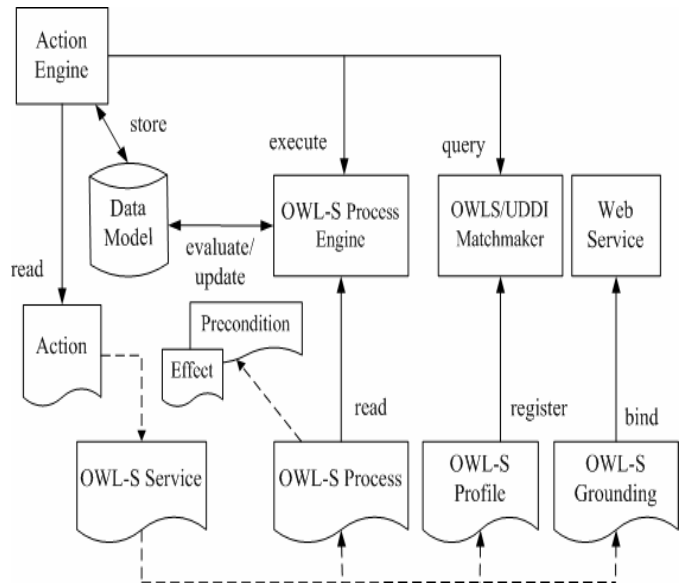

Figure 9: Agent Communication Framework for Semantic Web Services

On this extended architecture, we added: 1) Web service, 2) OWL/UDDI Matchmaker, and 3) OWL-Process Engine. We analyze the three components from the viewpoint of OWL-S: 1) they can be bound to a web service through OWL-S grounding, and they can be used by agents. 2) The OWL/UDDI Matchmaker provides the platform for the OWL-S Profile to register and match, 3) The OWL-Process Engine allows agent to invoke web services according to the OWL-S Process.

Our Action Engine is responsible for controlling these components. In reading action, it will store input into data model as parameters and use actions' capabilities to inquire OWLS/UDDI Matchmaker, attempting to find proper OWL-S service description, which will be read and executed by the

OWL process engine. Under three conditions: 1) before evaluation, 2) after updating precondition and effect and 3) after execution, the action engine will check the semantics of the conditions by using data model. The output of OWL-S will be stored into data model when 1) OWL-S process engine ends execution and 2) the action engine retrieves and sends the output back to the agent.

## V. AN EXAMPLE

We develop a movie recommendation system on the extended framework. In this system, there are four semantic web services: 1) Information Retrieval Service, 2) Recommendation Service, 3) Video Abstract Service and 4) Video Broadcast Service. Agents utilize cooperation and coordinate to interact with the semantic web service. With the cooperation, agent S queries the OWL-S/UDDI:

"Which agent(s) offer the abstract video service for "Brokeback Mountain"?"

Then, it receives the response:

"Agent T offers the service".

Then, agent S decides the CA to interact with agent T according to agent attitude. When it receives the request, it will judge FP of CA to response "agree" or "refuse". If agreed, it will execute the abstract video service. The process is shown in Figure 10.

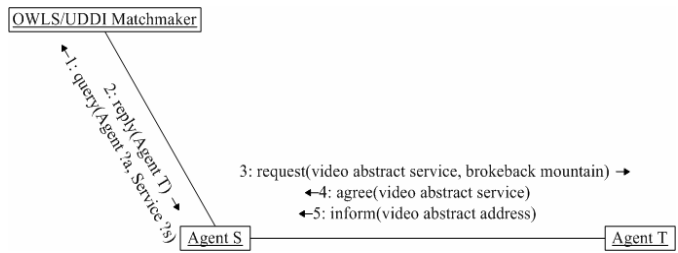

Figure 10: Cooperation with Semantic Web Service

When an agent queries OWL-S/UDDI and receives the response that involves multiple agents. The agent will use coordination to pick a proper agent. Assume agent S queries:

"Which agents provide video broadcast service to show the Brokeback Mountain?"

Then it receives the reply:

"The agent A, B, C, and D offer this service."

Then, it will use Contract Net Protocol to interact with them. Among them, agent c and d have difference condition bandwidth greater than 1Mbps and bandwidth greater than 2Mbps, respectively. Next, the agent S accepts the proposal of agent D. Finally, agent D will execute the video broadcast service. The process of coordination is shown in Figure 11, and the execution result of the movie recommendation system is shown in Figure 12.

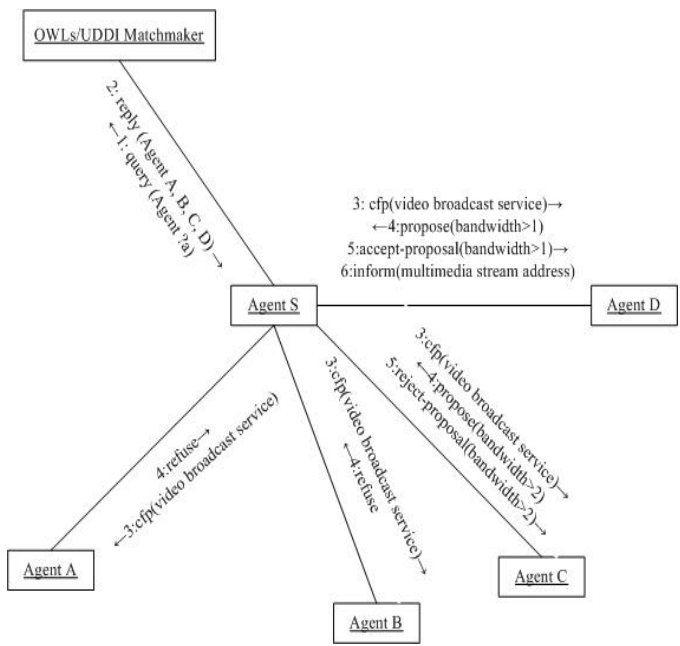

Figure 11: Movie Recommendation System

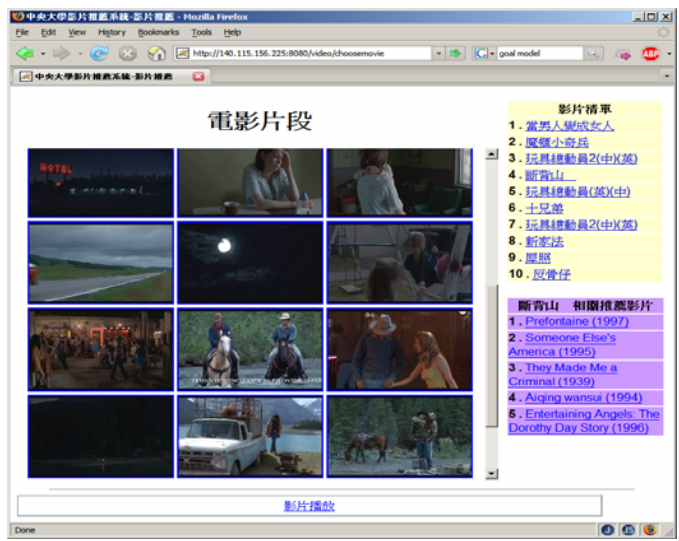

Figure 12: Movie Recommendation System

## VI. CONCLUSIONS

This paper defines the operational ontologies for agent communication, along with the engines to interpret the ontology. Further, through communication, agents can coordinate with each other to execute the semantic web services for users. This facilitates the declarative development of agent programs and share mental attitudes. Its advantages are as follow:

1) Dynamic Semantic Web Service Coordination: Operational ontology enables the calling agent communication to comprehend the meaning of the called. As a result, the calling agent communication can dynamically invoke the accurate semantic web service in accordance with semantics of operational ontology.

2) **Ontological Reasoning with Uniform Representation:** The operational ontology and domain ontologies stand for programs and data, respectively. They adopt the same representation format, namely declarative description. Finally, program and data on the semantic web can be inferred according to the declarative description.

3) **Easily Reuse and Extendable through Ontology Extension:** When a new ontology is needed, the ontology defined before can be extended by OWL. As for the new engines for the new ontology, we can also reuse the engines developed before.

## Acknowledgment

The authors wish to thank the anonymous reviewers for their valuable comments.

## AUTHORS PROFILE

**Chih-Hao Liu** received his Master degree of Information Engineering from Chaoyang University of Technology. He is currently a PhD candidate in the National Central University in Taiwan. He joined the software engineering laboratory in 2005. He also participated the SIM (Service-oriented Information Marketplace) project from 2005 to 2007. And, his current research interests focus on Semantic Web and Agent.

**Yong Feng Lin** received his PhD degree in Information Engineering and Computer Science from the National Central University, Taiwan, in 2007. His research interests are mainly on Semantic Web and Intelligent Agent. Recently, he is interested in applying modal extension of description logic to Semantic Web services.

**Jason Jen Yen Chen** is with the Department of Computer Science and Information Engineering in National Central University in Taiwan. He earned international recognition by winning Top, Third, and Fifth Scholar in the world in the field of System and Software Engineering in 1995, 1996, and 1997, respectively. The ranking is based on cumulative publication of six leading journals in that field. His current research interests include agile method and agent technology.